# Desublimation Frosting on Nanoengineered Surfaces

*Christopher Walker, Sebastian Lerch, Matthias Reininger, Hadi Eghlidi, Athanasios Milionis, Thomas M. Schutzius\*, Dimos Poulikakos\**

Laboratory of Thermodynamics in Emerging Technologies, Department of Mechanical and Process Engineering, ETH Zurich, Sonneggstrasse 3, CH-8092 Zurich, Switzerland

**Keywords: frosting, icing, condensation, desublimation, surface engineering, icephobic**

*To whom correspondence should be addressed.

Prof. Dimos Poulikakos
ETH Zurich
Laboratory of Thermodynamics in Emerging Technologies
Sonneggstrasse 3, ML J 36
CH-8092 Zurich
Switzerland
Phone: +41 44 632 27 38
Fax:+41 44 632 11 76
dpoulikakos@ethz.ch

Dr. Thomas M. Schutzius
ETH Zurich
Laboratory of Thermodynamics in Emerging Technologies
Sonneggstrasse 3, ML J 27.2
CH-8092 Zurich
Switzerland
Phone: +41 44 632 46 04
thomschu@ethz.ch

# Abstract

Ice nucleation from vapor presents a variety of challenges across a wide range of industries and applications including refrigeration, transportation, and energy generation. However, a rational comprehensive approach to fabricating intrinsically icephobic surfaces for frost formation—both from water condensation (followed by freezing) and in particular from desublimation (direct growth of ice crystals from vapor)—remains elusive. Here, guided by nucleation physics, we investigate the effect of material composition and surface texturing (atomically smooth to nanorough) on the nucleation and growth mechanism of frost for a range of conditions within the sublimation domain (0°C to -55°C; partial water vapor pressures 6 to 0.02 mbar). Surprisingly, we observe that on silicon at very cold temperatures—below the homogeneous ice solidification nucleation limit (< -46°C)—desublimation does not become the favorable pathway to frosting. Furthermore we show that surface nanoroughness makes frost formation on silicon more probable. We experimentally demonstrate at temperatures between -48°C and -55°C that nanotexture with radii of curvature within one order of magnitude of the critical radius of nucleation favors frost growth, facilitated by capillary condensation, consistent with Kelvin's Equation. Our findings show that such nanoscale surface morphology imposed by design to impart desired functionalities—such as superhydrophobicity—or from defects, can be highly detrimental for frost icephobicity at low temperatures and water vapor partial pressures (< 0.05 mbar). Our work contributes to the fundamental understanding of heterogeneous phase transitions well within the equilibrium sublimation domain and has implications for applications such as travel, power generation, and refrigeration.

Over the last decade, research interest on nucleation physics and the rational design of icephobic surfaces has been experiencing rapid growth due to exciting recent findings[1–6] and the ability to probe and explain previously unexplored phenomena for example distinct frost growth patterns on differently chemically treated surfaces that allow easier frost removal[2] or the spontaneous jumping of droplets coalescing on hydrophobic surfaces to prevent freezing of supercooled droplets.[4,7,8] Going beyond scientific value, such research is crucial in understanding how to engineer surfaces for commercial usage to benefit practical applications.[9–12] To this end, industries spanning from road and air transportation, to transmission lines, to energy generation by wind turbines and to refrigeration cycles[13,14] can harvest significant benefit from the development of icephobic surfaces with better performance under temperatures far below the equilibrium freezing point. Icephobicity at the very fundamental ice formation level, can be characterized as the ability to hinder heterogeneous ice nucleation on the surface.[15] An example of heterogeneous ice nucleation, commonly referred to as frosting, is the formation of ice crystals on a surface, at a temperature below the freezing point and exposed to a supersaturated vapor environment.[16] This ice nucleation can follow two pathways: vapor to liquid to solid (condensation-freezing) or vapor to solid (deposition or desublimation).[17] These two pathways, in a variety of forms, have long been an area of interest in the atmospheric sciences due to their importance for understanding how particles in the environment influence the formation of clouds, fog, rain, sleet, and snow,[18] and this work has focused on either inorganic crystals or natural particles commonly found in the atmosphere that expedite ice nucleation.[19–23] Understanding what makes a surface icephobic, however, must involve the intertwined effects of surface material types and properties with respect to counteracting ice nucleation. Furthermore, understanding how these factors affect the mechanism and transition between the two pathways of heterogeneous ice nucleation from vapor is crucial in understanding how to design and tailor icephobic surfaces for different applications.[24]

The effect of surface topography on ice nucleation has been a topic that has received renewed attention in recent years, due to the increased interest in developing inherently icephobic surfaces and the increasing ability to impart topography with desired nanoscale features in a controlled manner. However, due to the number of nucleation pathways,[17] there still remain considerable aspects not completely understood, including the performance and pathway dependence under a range of environmental conditions. For example, with solidification, studies have concluded that nanotexture, with radii of curvature within one order of magnitude of the critical radius, does not facilitate nucleation of super-cooled water droplets (heterogeneous nucleation from the melt), although classical nucleation theory would predict it.[25,26] These results are explained either by the formation of an amorphous quasi-liquid layer at the interface between the nascent ice embryo and the surface, counteracting the decreased interfacial energy which should otherwise be provided by the pits[25] or due to the relatively higher interfacial energy between the liquid and solid in comparison to that of the liquid and vapor, considerably dampening the geometrical effect of the pits to enhance nucleation rate.[26] Evidence that nanotexture and nanoscale surface defects improve crystal nucleation from vapor have been previously shown, such as chemical vapor deposition of diamond[27,28] as well as the growth of organic crystals.[29,30] However, despite its importance, the effect of nanotexture on water vapor desublimation, at the intersection of the science bases of ice nucleation and surface nanofabrication, remains largely unexplored.

Here, employing classical nucleation theory in the sublimation domain of the water equilibrium diagram, we begin by mapping the expected homogeneous nucleation mode as a function of temperature and super-saturation and discuss how the introduction of a surface may affect the nucleation behavior. First, we verify our experimental apparatus and methodology by comparing our own findings on silver iodide crystals to those in literature. We use these experiments to begin our discussion on how substrate material affects the barrier to heterogeneous ice nucleation and the heterogeneous condensation-desublimation nucleation transition. We then investigate silicon and the effect of surface morphology, namely, nanotexturing, on frost nucleation behavior. Based on our experimental findings, we discuss how ice nucleates from vapor under different conditions on a host of substrate materials and surface topographies, and suggest rational design rules that must be considered when engineering icephobic surface textures for realistic applications.

## Results and Discussion

The mechanism of nucleation is characterized by the change in Gibbs free energy necessary to transform from a parent phase (1) to a daughter phase (2).[17,18] In frosting, the parent phase is always water vapor, while the daughter phase can be either liquid water (condensation) or solid ice (desublimation), dependent on the nucleation pathway; in the case of liquid formation, this will then subsequently solidify. Since our goal is simply to identify the conditions necessary for the two pathways, we have ignored the Gibbs free energy necessary for the liquid to freeze in the ensuing analysis. The critical Gibbs free energy ($\Delta G_{12}^*$) indicates the amount of energy necessary to begin phase change from the parent phase (1) to the daughter phase (2) at a specific temperature ($T$) and pressure ($p$) (single component). $\Delta G_{12}^*$ for each nucleation pathway described above, is given as

$$\Delta G_{VL}^* = 16\pi \gamma_{VL}^3 \xi / 3 \left[ n_L kT \ln(p/p_L) \right]^2 \text{, and} \tag{1}$$

$$\Delta G_{VS}^* = 16\pi \gamma_{VS}^3 \xi / 3 \left[ n_S kT \ln(p/p_S) \right]^2, \tag{2}$$

(see Supporting Information, section "Critical Gibbs Free Energy") where $\gamma$ is the interfacial energy between the parent and daughter phases, $n$ is the number of molecules per unit volume of the daughter phase, $k$ is the Boltzmann constant, $p_L$ and $p_S$ are the saturated water vapor pressures over a plane surface of the daughter phase, and $\xi$ is a numerical factor greater than one used to compensate for non-spherical embryos (for the vapor to liquid transition $\xi$ is equal to one). The subscripts $V$, $L$, and $S$ take the place of parent phase (1) and daughter phase (2) explicitly as the vapor phase, liquid phase, and solid phase, respectively.

An important parameter to study, dependent on $\Delta G_{12}^*$, is the nucleation rate of the daughter phase ($J_{12}$). The rate at which nucleation by condensation ($J_{VL}$) and desublimation ($J_{VS}$) occurs is calculated by

$$J_{VL} = J \exp\left(-\Delta G_{VL}^*/kT\right), \text{ and} \tag{3}$$

$$J_{VS} = J \exp\left(-\Delta G_{VS}^*/kT\right), \tag{4}$$

where $J$ is a pre-exponential term, dependent on the rate at which vapor molecules strike a unit area of the embryo, and is the same for condensation and desublimation (see Supporting Information, section "Nucleation Rate"). Consequently, as we will also further elucidate below, if $\Delta G_{VS}^* < \Delta G_{VL}^*$ the nucleation rate of solid ice crystals (desublimation) will dominate the nucleation rate of liquid droplets ($J_{VS} > J_{VL}$).

Figure 1a-b plots the change in free energy of a forming liquid ($\Delta G_{VL}$) and solid water embryos ($\Delta G_{VS}$) from vapor vs. embryo radius, $r$, for two different environmental conditions (see Supporting Information, section "Critical Gibbs Free Energy" equation S1). In Figure 1a, $T$ = 225 K and $p$ = 0.3 mbar, and in Figure 1b, $T$ = 215 K and $p$ = 0.05 mbar; the latter case is colder and drier relative to the former case. It is clear in Figure 1a and b that $\Delta G_{VL}^* - \Delta G_{VS}^* < 0$ and $\Delta G_{VL}^* - \Delta G_{VS}^* > 0$, respectively; therefore, in the former case condensation is preferred while in the latter case desublimation nucleation is preferred. This analysis indicates that the relative magnitudes of $\Delta G_{VS}^*$ and $\Delta G_{VL}^*$ can switch based upon differing environmental conditions. For example, in this case, decreasing $T$ and $p$ causes a change from condensation to desublimation nucleation. We note that this transition occurs at temperatures far from the solidification temperature and for very dry conditions. Figure 1c plots $p/p_S$ vs. $T$ vs. $\Delta G_{VL}^* - \Delta G_{VS}^*$ and quantifies the conditions where a liquid or solid daughter phase is favored. It is clear that at relatively low $T$ and $p/p_S$,

$(\Delta G_{VL}^* - \Delta G_{VS}^*)/kT > 0$, and desublimation is preferred (Eq.(2)). At higher $T$ and $p/p_S$, condensation is favorable. Due to the sub-freezing temperatures and the corresponding metastable state some of the condensate will eventually freeze, becoming frost (condensation freezing).

The above discussion of homogeneous nucleation is the first step in understanding heterogeneous nucleation, which is nucleation in the presence of a foreign substance. It is ubiquitous in almost all related applications due to the fact $J_{VL}$ and $J_{VS}$ only become appreciable when nucleation occurs heterogeneously.[22] When considering heterogeneous nucleation, $\Delta G_{VL}^*$ and $\Delta G_{VS}^*$ are modified by a function, $f(\theta_{12}, R) \leq 1$, which depends on the inherent contact angle of the daughter phase on the surface in the parent phase ($\theta_{12}$) and the radius of curvature of the foreign body ($R$) to give us the heterogeneous critical change in Gibbs free energy: $\Delta G_{12}^* f(\theta_{12}, R)$ ( see Supporting Information, section "Homogeneous and Heterogeneous Nucleation: Geometric Factor"). A surface therefore can change the condensation-desublimation nucleation transition dependent on the surface properties, namely the intrinsic contact angle of water ($\theta_{VL}$), the intrinsic contact angle of ice ($\theta_{VS}$), and $R$, but also considerably lowers the magnitude of $\Delta G_{12}^*$, resulting in a drastic increase in $J_{12}$. To underpin this, Figure 1d plots $p/p_S$ vs. $T$ vs. $J_{12}$ using Eq. (3)-(4) for the same range of $T$ (200 – 273 K) and $p/p_S$ (1.35 – 10) for a heterogeneous nucleation case, where we choose values of $\theta_{VL} = \theta_{VS} = 75°$ and $R = \infty$ to most simply and effectively illustrate the influence of a foreign surface. Eq. (4) is used when desublimation nucleation is favorable (left of solid black line) and Eq. (3) is used when condensation nucleation is favorable (right of solid black line). The orange line represents $J_{12}$ = 1 embryo cm$^{-2}$ s$^{-1}$ and is considered a lower limit of an appreciable nucleation rate.[31] The value of $\theta_{VS}$ is a function of material composition, which from the above discussion clearly shows its importance on affecting nucleation along with $R$. Since $\theta_{VS}$ cannot be defined *a priori*—along with the impact of $R$ on it—the heterogeneous nucleation behavior, namely the pathway to nucleation on any surface at different substrate temperature ($T_s$) and $p$ must be experimentally investigated. The experimental determination of $T_s$ at which the nucleation pathway changes and $p/p_S$ in the desublimation regime, however, offers a solution to experimentally measure the apparent ice-vapor-substrate contact angle, $\theta_{VS}^*$, similar to how conventional goniometers measure the apparent water contact angle, $\theta_{VL}^*$ (see Supporting Information, section "Ice-vapor-substrate Contact Angle").

In order to investigate the heterogeneous condensation-desublimation nucleation transition, we examined condensation and desublimation behavior on silver iodide coated surfaces (AgI), a well-known nucleation agent.[32] Figure 2a shows a plot of $p/p_S$ vs. $T_s$ vs. nucleation mode on AgI; we denote two regions: $p > p_L$ (refers to region of $p/p_S > p_L/p_S$) and $p < p_L$ (refers to region of $p/p_S < p_L/p_S$). When $p > p_L$, the vapor is super-saturated with respect to a flat liquid surface. When $p > p_S$ ($p/p_S > 1$), the vapor is supersaturated with respect to a flat solid surface. Due to the limitations set by the size of $r^*$ (of the order of nm) it is impossible to observe the nascent embryo at the onset of nucleation in our experimental setup; therefore, we attributed a condensed nucleating phase for $p < p_L$ and $p > p_S$ to desublimation. (This is further strongly supported by observing that the growing ice crystal is a polyhedron; *cf.* Figure 2c.) We observed that for relatively higher values of $T_s$ (> -12 °C), $p > p_L$ when we observe nucleation and the nucleate consists of metastable liquid water, which given sufficient time freezes, Figure 2b (condensation regime). When $T_s$ < -12 °C, the onset of nucleation takes place at $p < p_L$ and $p/p_S > 1$, and we observe the growth of an ice crystal with six-fold symmetry, Figure 2c. Additionally, Figure 2a compares our results (purple data points) with those obtained in previous studies[19,21] (dark grey and light grey data points). Their good agreement in heterogeneous condensation-desublimation nucleation transition $T_s$, value of $p/p_S$ in the desublimation regime, and the resulting calculation of $\theta_{VL}^*$ and $\theta_{VS}^*$ is used to confidently confirm the accuracy of our experimental apparatus and methods (see Supporting Information, section "Wettability of AgI").

Due to its high affinity to ice, AgI proved to be a good substrate to examine the role that substrate material plays in altering $\Delta G_{VL}^*$ and $\Delta G_{VS}^*$ at relatively high $T_s$ (> -20°C). However, the inherent inhomogeneous surface texture created when fabricating the AgI samples, seen in Figure 2e, made it a poor candidate to decouple the effects of $\theta_{12}$ and $R$ in order to study the effect of nanotexture on the nucleation behavior from vapor. We chose silicon as a representative substrate to fundamentally investigate the effect of surface roughness on desublimation nucleation due to its ability to be engineered to specific values of $R$ and the existence of reliable methodologies to impart corresponding desired surface textures.

Classical nucleation theory indicates that roughness, whose radius of curvature $R$ is up to one order of magnitude above the minimal radius of curvature of stable nuclei ($r_{12}^*$), should affect the nucleation energy barrier of the daughter phase at the same conditions as a surface with no roughness. Figure 3a and b show the theoretical influence that nanotexture has on $f(\theta_{12}, R)$, within the experimental range of $p$ and $T_s$ that were investigated. Due to difficulty in exactly determining $\theta_{VS}$ and $\theta_{VL}$ for silicon, we show the influence for three different plausible $\theta_{12}$ values that we calculate, which correspond well to previously calculated ice and liquid contact angles values for silicon.[25] (see Supporting Information, section "Wettability of Silicon"). Examining both Figure 3a and b it is clear that, independent of $\theta_{12}$, smaller $R$ values have larger influence on $f(\theta_{12}, R)$. Due to small convex (positive) values of $R$ (< 100 nm), $f(\theta_{12}, R)$ increases (Figure 3a), leading to an increase in $\Delta G_{12}^* f(\theta_{12}, R)$ and hence a decrease in $J_{12}$ (Eq.(3) and (4)). Due to small concave (negative) values of $R$ (< 100 nm), the opposite effect happens, namely $J_{12}$ increases. Figure 3c and d show micrographs obtained with atomic force microscopy using a 2-nm extra sharp tip, overlaid with a surface plot of mean curvature ($H = 1/R$) for our smooth (untreated wafers, RMS roughness

= 0.94 nm) and nanotextured silicon (RMS roughness = 61.8 nm) samples, respectively. The histograms shown in Figure 3e and f indicate the percent area fraction of the surface covered by concave pits of radius $R$ (negative values) for our smooth and nanotextured silicon samples, respectively. Comparing the smooth and nanotextured silicon surfaces, we observed that the nanotextured surface has 18% of its area—a significant fraction—covered with nanopits with $R < 10\, r_{12}^*$, while the smooth silicon has effectively none of its area covered with nanopits with $R < 10\, r_{12}^*$.

Figure 3 provides evidence that the nanotextured surface shown in d has both bumps and pits that are on a scale capable of drastically changing $f(\theta_{12}, R)$ (as indicated by the grey region in a and b, matching magnitude of $R$ in the grey region in f). Since both surface curvatures are inherently created when fabricating roughness—where bumps and pits inhibit or promote nucleation respectively—it is unclear at first glance what the effect of nanotexture will be, with respect to the nucleation rate. By performing an analytical comparison of $J_{12}$ for a smooth surface and a nanotextured surface, it becomes clear that a nanotexture surface greatly increases $J_{12}$, due to the dominance of the pits (see Supporting Information, section "Effect of Roughness on Nucleation Rate").

We tested the effect of nanotexture on the nucleation mechanism at these conditions by conducting experiments on smooth silicon and on nanotextured silicon (see Methods, section "Surface Preparation" for more details) at four values of $p$. The sample was observed using a bright field microscope, while its temperature ($T_s$) was slowly reduced. Upon first observation of nucleation, $T_s$ was recorded and used to calculate $p_S$ (see Methods, section "Setup and Experimental Protocol" for more details). Figure 4a shows a box plot of super-saturation, $p/p_S$ at the moment when nucleation was observed, vs. $p$ for smooth and nanotextured silicon samples; each box includes a total of nine experiments on each sample type using three different samples of each type. The values $p$ of 0.05, 0.04, 0.03, and 0.02 mbar used in the experiments correspond to equilibrium vapor pressures over flat ice at temperatures ($T_\infty$) of -47.9, -49.7, -52.0, and -55.2 °C, respectively. Examining the plot it is clear that $p/p_S$ is considerably lower for the nanotextured samples, implying that $J_{12}$ has been considerably increased. We interpret our experimental results to indicate that $R$ of the order of magnitude of $r_{12}^*$, does indeed improve the favorability of ice growth in our experimental temperature range. In order to further confirm the influence on nanotexture on nucleation rate, and its importance in technical applications, we conducted experiments on three nanotextured aluminum samples at a $p$ value of 0.05 mbar (9 total experiments). Similar to nanotextured silicon, we observed that these surfaces resulted in an increased $J_{12}$ at reduced $p/p_S$ and therefore an earlier onset of nucleation, seen by the red box in Figure 4a. This provides stronger evidence for the effect of nanotexture across a range of materials to increase $J_{12}$.

It is also interesting to observe the shape of the ice growing on the silicon surface. Figure 4b and c show images of ice growing on smooth and nanotextured silicon, respectively. It is clear that the ice growing on both surfaces is not symmetric like it was on AgI coated surfaces, although the cold temperature and dry environment would lead one to expect to be in a desublimation regime. Therefore, it is unclear that desublimation nucleation is the favorable pathway to ice formation. No clear condensation-desublimation nucleation transition—as we were able to obtain for AgI coated surfaces (Figure 2a)—was attained on smooth silicon for the temperature ranges we studied (see Figure S1a). Furthermore, the macroscopic shape of the ice crystals provides insight into the microscope of crystal configuration upon nucleation and therefore evidence for the interface at which growth began, whether it be solid-vapor or solid-liquid.[33] The basal face of ice 1h crystals observed on the AgI samples (Figure 3c) is consistent with growth beginning from the solid-vapor interface, while the secondary prism face of ice 1h crystals observed on the silicon samples (Figure S1b) is indicative of growth from the solid-liquid interface,[34] providing further evidence that condensation followed by freezing is the energetically favorable pathway for frost formation for the range of $T_s$ used in our experiments. Nevertheless, our observations show that nanotexture does influence nucleation and catalyzes the formation of ice. We offer two feasible explanations that we see for this behavior: 1) The nanoroughness radius of curvature $R$ of the scale of $r_{VS}^*$ is able to significantly reduce $f(\theta_{VS}, R)$ ( see Supporting Information, section "Homogeneous and Heterogeneous Nucleation: Geometric Factor") and ice desublimation is thermodynamically favorable at lower $p/p_S$ or 2) the extreme temperatures in these experiments (< -48° C), makes it likely for a miniscule volume of metastable liquid formed in the nanotexture through capillary condensation to freeze, enabling the pathway to frost nucleation at lower $p/p_S$.

Due to the crystalline structure of ice, it is expected that adding curvature at such a small scale will disrupt the lattice near the substrate surface, resulting in an amorphous phase known as a quasiliquid layer, whose thickness increases with decreasing pit radius,[25,35] contradicting our first explanation. The thickness or actual existence of the quasiliquid layer may be dependent on substrate type, including crystallographic match with respect to bulk ice and the strength of the water—surface interaction.[36] The idea that ice crystal growth can be enhanced by surface defects as a two-step process involving a liquid intermediary, however, has been proposed earlier[37–40] and has been experimentally demonstrated for a variety of organic liquids.[41] The proposed mechanism for the enhanced growth of ice crystals from vapor involves an intermediary liquid phase cause by capillary condensation. The phenomenon of capillary condensation can be described by the Kelvin Equation,

$$\ln(p/p_L) = \frac{-4\gamma_{VL} \cos\theta_{VL}}{n_L dkT}, \tag{5}$$

where $d$ is the diameter of a cavity. Eq. (5) illustrates that if $\theta_{VL}$ < 90°, then the equilibrium vapor pressure over the interface ($p$) can be smaller than that of the bulk equilibrium vapor pressure ($p_L$) and result in stable condensate that forms at under-saturated vapor pressures. Even if the $\theta_{VL}$ > 90°, capillary condensation can take place at considerably smaller $p$ than in absence of nanotexture. Although, miniscule volumes of condensate can form inside these cavities, emergence is not thermodynamically favorable for the liquid condensate at under-saturated vapor pressure with respect to liquid water, which limits any macroscopic growth of the condensate until the appropriate supersaturated conditions have been reached.[30] However, very low temperatures considerably increase the ice embryo nucleation rate, rendering it still sufficient for nucleation events to occur in a reasonable time, even in the miniscule liquid volumes associated with capillary condensate in our nanotexture. In contrast to the metastable liquid condensate, the now-nucleated ice crystals are able to grow out of the nanotexture due to the super-saturation above the thermodynamically stable ice crystals.[41]

This explanation also warrants a discussion on the effect of freezing point depression (Gibbs-Thompson Effect) on the nucleation of ice embryos in the nanotexture. It has been shown that water inside nanopores with d < ~12nm leads to lower equilibrium solidification temperatures and complete suppression of solidification for radii under 2.8nm.[35,42,43] This effect, akin to the formation of quasiliquid layers at locations of high $R$,[25,35] should, in theory, only suppress ice nucleation at the very bottom of the nanopits where the pit diameter is of the correct scale. The nanotexture on the tested surfaces are characterized by an RMS roughness of 61.8nm and have peak-to-peak distances on the order of 100nm, implying that if the capillary condensate is able to fill the nanopits ($p/p_L > 1$), then the ice nucleation should still be able to nucleate homogeneously inside the condensate, at the side-wall surface interface, or at the three-phase line. It is worth mentioning that the effect of the low $p$ used in our experiments may have an influence on the adsorption of water molecules on the surface and thereby may influence the surface energies of liquid-vapor and solid-vapor;[44] however, the situation is complicated by the presence of other adsorbates such as volatile organic compounds, which previously has been shown to increase surface hydrophobicity.[45]

Based upon this discussion and our experimental results, we conclude that metastable capillary condensate forming within cavities created by the nanotexture and its subsequent freezing at low temperatures, allows ice crystals to form and grow at lower $p/p_S$ than on the smooth silicon samples, due to the lack of "seed" condensate on smooth samples. An analysis of this discussion and comparison between the nanotexture and smooth surfaces is illustrated in Figure 4d. In this nucleation pathway we note that the capillary condensate only plays a role in initiating frost growth, when significantly low $T_s$ are present to enable a substantial nucleation rate of ice for the small capillary condensate volumes. At higher $T_s$ the capillary condensate should also be present, but does not have the chance to propagate out of the nanotexture, due to the resulting low $J_{LS}$.[46]

## Conclusions

In this work we revisit the analysis of the heterogeneous condensation-desublimation nucleation transition. Our motivation is to understand it in the context of surface icephobicity in order to advise the engineering of rationally designed icephobic surfaces under low water vapor pressure and low temperature conditions. After rigorously defining and experimentally validating the regions of nucleation behavior as condensation or desublimation based upon $T_s$ and $p$ for a well-known ice nucleation agent, we study a nanotextured substrate to quantitatively define the effect of surface nanotopography on the barrier to frost nucleation and define a mechanism for nucleation behavior at low temperatures.

Our results are significant as they build on our understanding of the nucleation process of ice crystals from vapor. We observe that true desublimation can indeed occur on specific materials, whose solid-substrate interfacial energy is equal to or less than its liquid-substrate interfacial energy. Although these materials may be of high interest in the atmospheric sciences, their use in the design of icephobic surfaces is less obvious. On surfaces relevant for icephobicity, we surprisingly found that the nucleation pathway at low temperatures on sufficiently nanotextured surfaces (R < 25 nm, one order of magnitude above the minimal radius of curvature of stable ice nuclei, $r_{12}^*$) differs from the expected desublimation pathway. The substantially low temperatures necessary to achieve desublimation nucleation on such surfaces, enable surface topography of the right scale to trigger a different pathway to crystal growth, namely freezing of metastable capillary condensate, resulting in macroscopic crystal growth at lower super-saturation.

To this end, we conclude that the design of icephobic surfaces for very low-temperature applications place a significant amount of effort to use surfaces with as few nanoscale defects as possible to avoid ice crystal growth through homogeneous freezing of capillary condensate. Nonetheless, it has recently been reported that nanotexture and roughness decreases frost adhesion,[47,48] rendering the question of whether or not nanoengineered sample texture should be used, highly dependent on the area of application.

In summary, we find that smoother surfaces outperform nanotextured surfaces under low temperature and vapor pressure conditions. We conclude that most engineering relevant surfaces (silicon, metal, plastic) will exhibit this behavior, as long as their lattice spacing and chemical composition do not render it a strong ice nucleator (such as AgI), which has very important implications for applications such as air travel, power generation and condenser technology.

# Methods

Surface Preparation

The following cleaning protocol was used at different steps during the surface fabrication: four-minute sonication in baths of acetone, isopropanol, and water followed by a two-minute 600W oxygen plasma (PVA TePla, GIGAbatch 310M). Silver iodide (AgI) samples were made on 175 µm thick, one-inch diameter glass cover slides. The glass slides were initially cleaned using described cleaning protocol and then coated with a 2 nm layer of evaporated titanium (to improve silver adhesion), followed by a 10 nm layer of evaporated silver (Evatec, BAK501). The coated glass slides were then placed in an air-tight environment for 60 seconds containing a beaker of 2% vol. iodine (Sigma-Aldrich, ≥99.99% trace metals basis) in deionized water solution which was heated to 60 °C. The iodide vapor reacted with the silver coating to form AgI crystals, confirmed by change in sample color and SEM micrographs (refer to Figure 2d and e). Silicon samples were fabricated from four-inch, 100 μm thick, double-side polished, p-doped silicon wafers with <100> crystal orientation. The wafers were diced into 1x1 cm chips. All silicon samples were initially cleaned using the described cleaning protocol. After cleaning, the smooth silicon samples were used as is. We fabricated the nanotextured silicon samples by first sputtering a 2 nm layer of gold onto the silicon surface (VON ARDENNE GmbH, Cluster System CS3205), which resulted in a nanoscale inhomogeneous gold layer. The gold coated silicon chips were exposed to a 50W reactive ion etching process using 100 sccm flow rate sulfur hexaflouride, 40 sccm flow rate oxygen at 20°C sample temperature and 40 mTorr chamber pressure for two minutes (Oxford Instruments, Plasmalab 80 Plus). Nanotextured aluminum samples were fabricated by immersing 100 μm thick aluminum sheets (Korff AG, Switzerland) in boiling deionized water. Prior to immersion the aluminum samples were cleaned by sonication in acetone, isopropyl alcohol, and deionized water baths (10 minutes for each solvent). After immersing the aluminum in the boiling water, the native oxide layer on the surface reacts and becomes hydrolyzed. The oxide film undergoes chemical and morphological change leading to the formation of boehmite nanostructures as can be seen by the images of Figure S2.

Surface Characterization

The AgI and aluminum samples were characterized by performing scanning electron microscopy (Hitachi, SU8000) scans. The silicon nanotextured samples were characterized by performing atomic force microscopy (Bruker, Dimension FastScan) scans in tapping mode using 2 nm sharp tips.

## Setup and Experimental Protocol

All experiments were performed using a gas-tight chamber illustrated in Figure 5, capable of reaching absolute pressures ($p_0$) < 0.2 mbar with the use of a roughing vacuum pump (Vacuubrand GmbH, RE 2.5 rotary vane pump). During experiments it was imperative we obtain the highest possible accuracy of $p$, due to its important influence on $\Delta G_{12}^*$ and ultimately $J_{12}$. We were most accurately able to measure and precisely control $p$ by altering the temperature ($T_\infty$) of a copper block, on which a thin layer of source ice was grown. Relating the temperature of the thin ice layer on the block (assumed to be the same as the copper block) to the equilibrium vapor pressure over a flat interface of ice[49] resulted in the calculation for $p$. Furthermore, the source ice block was designed to be 100 times the area of the sample ice nucleation area in order to maintain a constant $p$ during nucleation and growth of the ice on the sample. By allowing enough time for the source ice to equilibrate with the water vapor in the chamber, a high accuracy for $p$ could be achieved. We assumed that the $p$ in the chamber had reached equilibrium when the difference between the temperature measurement inside the copper block ($T_\infty$) and on the copper block ($T_b$) became constant, indicating no further change in heat flux through the source copper block. The sample temperature ($T_s$) was controlled by regulating the temperature of a copper block in contact with the sample. $T_s$ and $T_\infty$ could be independently regulated using the combination of two independent cold nitrogen vapor lines and two independently PID-controlled thermoelectric elements in contact with each of the respective copper blocks. $T_s$ and $T_\infty$ were measured using resistance temperature detectors (IST AG, Pt1000, +/- 0.15°C) located inside each respective copper block, approximately 1 mm behind the exposed surface. A viewing window at the bottom of the chamber enabled the observation of the sample during the nucleation process using a home-built inverted microscope (see Figure S3 for details on optical train). A CMOS USB camera (Thorlabs, DCC1645C), data acquisition system (Beckhoff, Ethercat EK1100), and signal processing software (National Instruments, LabVIEW 2015) enabled the synchronized recording of the images and temperatures during experiments for later analysis.

At the beginning of each set of experiments, the chamber was pumped down and held at < 0.2 mbar until the desired $T_\infty$ was reached. Water vapor was then allowed to enter the chamber, to set $p$, either by opening a valve exposing the chamber to a deionized water vapor bath (AgI experiments) or exposing the chamber to atmospheric lab conditions by opening the vacuum release valve (silicon experiments). The latter was done for the silicon experiments due to the low $T_\infty$ used and the difficulty it posed in maintaining a vacuum-tight (and hence a water vapor tight) chamber. After the introduction of water vapor we immediately observed the formation of ice on the source copper block based upon the observation of an isothermal increase in the signal measuring $T_\infty$. We then slowly decreased $T_s$ at a rate of < 0.1 °C/s. At some $T_s < T_\infty$ nucleation was observed and $T_s$ at the first onset of nucleation was recorded, allowing us to calculate $p_S$, and therefore the super-saturation ($p/p_S$) at first onset of nucleation.

## Supporting Information

**Supporting Information Available:** The supplemental information contains further information on Critical Gibbs Free Energy, Nucleation Rate, Homogeneous and Heterogeneous Nucleation: Geometric Factor, Ice-vapor-substrate Contact Angle, Wettability of AgI, Wettability of Silicon, and Effect of Roughness on Nucleation Rate. Additionally, the following figures are included: Experimental nucleation conditions for silicon, SEM micrograph of boehmite nanostructures on aluminum, and Optical train used for nucleation observations. Video S1: Condensation vs. Desublimation nucleation on AgI. This material is free of charge *via* the Internet at http://pubs.acs.org.

## Acknowledgements

We thank J. Vidic and P. Feusi for assistance in experimental setup construction. We thank U. Drechsler for assistance in surface engineering. Partial support by the Swiss National Science Foundation under Grant 162565 and the European Research Council under Advanced Grant 669908 (INTICE) is acknowledged. T.M.S. also acknowledges the ETH Zurich Postdoctoral Fellowship Program and the Marie Curie Actions for People COFUND Programme (FEL-14 13-1).

## Author Contribution Statement

T.M.S. and D.P. designed research and provided scientific advice on all its aspects; C.W., S.L., M.R., and H.E. implemented experiments; C.W. and A.M. fabricated and characterized samples; C.W. and S.L. performed research and analyzed data; and C.W., T.M.S., and D.P wrote the paper.

## Additional Information

**Competing financial interests:** The authors declare no competing financial interests.

# References

(1) Varanasi, K. K.; Deng, T.; Smith, J. D.; Hsu, M.; Bhate, N. Frost Formation and Ice Adhesion on Superhydrophobic Surfaces. *Appl. Phys. Lett.* **2010**, *97*, 234102.

(2) Liu, J.; Zhu, C.; Liu, K.; Jiang, Y.; Song, Y.; Francisco, J. S.; Zeng, X. C.; Wang, J. Distinct Ice Patterns on Solid Surfaces with Various Wettabilities. *Proc. Natl. Acad. Sci.* **2017**, *114*, 11285-11290.

(3) Graeber, G.; Schutzius, T. M.; Eghlidi, H.; Poulikakos, D. Spontaneous Self-Dislodging of Freezing Water Droplets and the Role of Wettability. *Proc. Natl. Acad. Sci.* **2017**, *114*, 11040–11045.

(4) Boreyko, J. B.; Collier, C. P. Delayed Frost Growth on Jumping-Drop Superhydrophobic Surfaces. *ACS Nano* **2013**, *7*, 1618–1627.

(5) Rykaczewski, K.; Anand, S.; Subramanyam, S. B.; Varanasi, K. K. Mechanism of Frost Formation on Lubricant-Impregnated Surfaces. *Langmuir* **2013**, *29*, 5230–5238.

(6) Kajiya, T.; Schellenberger, F.; Papadopoulos, P.; Vollmer, D.; Butt, H. J. 3D Imaging of Water-Drop Condensation on Hydrophobic and Hydrophilic Lubricant-Impregnated Surfaces. *Sci. Rep.* **2016**, *6*, 23687.

(7) Zhang, Q.; He, M.; Chen, J.; Wang, J.; Song, Y.; Jiang, L. Anti-Icing Surfaces Based on Enhanced Self-Propelled Jumping of Condensed Water Microdroplets. *Chem. Commun.* **2013**, *49*, 4516–4518.

(8) Liu, J.; Guo, H.; Zhang, B.; Qiao, S.; Shao, M.; Zhang, X.; Feng, X. Q.; Li, Q.; Song, Y.; Jiang, L.; Wang, J. Guided Self-Propelled Leaping of Droplets on a Micro-Anisotropic Superhydrophobic Surface. *Angew. Chemie - Int. Ed.* **2016**, *55*, 4265–4269.

(9) Boreyko, J. B.; Hansen, R. R.; Murphy, K. R.; Nath, S.; Retterer, S. T.; Collier, C. P. Controlling Condensation and Frost Growth with Chemical Micropatterns. *Sci. Rep.* **2016**, *6*, 19131.

(10) Sun, X.; Rykaczewski, K. Suppression of Frost Nucleation Achieved Using the Nanoengineered Integral Humidity Sink Effect. *ACS Nano* **2017**, *11*, 906–917.

(11) Lo, C. W.; Sahoo, V.; Lu, M. C. Control of Ice Formation. *ACS Nano* **2017**, *11*, 2665–2674.

(12) Verho, T.; Bower, C.; Andrew, P.; Franssila, S.; Ikkala, O.; Ras, R. H. A. Mechanically Durable Superhydrophobic Surfaces. *Adv. Mater.* **2011**, *23*, 673–678.

(13) Schutzius, T. M.; Jung, S.; Maitra, T.; Eberle, P.; Antonini, C.; Stamatopoulos, C.; Poulikakos, D. Physics of Icing and Rational Design of Surfaces with Extraordinary Icephobicity. *Langmuir* **2015**, *31*, 4807–4821.

(14) Lv, J.; Song, Y.; Jiang, L.; Wang, J. Bio-Inspired Strategies for Anti-Icing. *ACS Nano* **2014**, *8*, 3152–3169.

(15) Kreder, M. J.; Alvarenga, J.; Kim, P.; Aizenberg, J. Design of Anti-Icing Surfaces: Smooth, Textured or Slippery? *Nat. Rev. Mater.* **2016**, *1*, 15003.

(16) Nath, S.; Ahmadi, S. F.; Boreyko, J. B. A Review of Condensation Frosting. *Nanoscale Microscale Thermophys. Eng.* **2017**, *21*, 81–101.

(17) Pruppacher, H. R.; Klett, J. D. Microphysics of Clouds and Precipitation. In *Atmospheric and Oceanographic Sciences Library*; Mysak, L.A., Hamilton K., Eds.; Springer: Heidelberg, 1998.

(18) Hobbs, P. V. Ice Physics. Oxford University Press: Oxford, 1974; pp. 461–571.

(19) Bryant, G. W.; Hallett, J.; Mason, B. J. The Epitaxial Growth of Ice on Single-Crystalline Substrates. *J. Phys. Chem. Solids* **1960**, *12*, 189–195.

(20) Layton, R. G.; Steger, J. Nucleation of Ice on Silver Iodide. *J. Atmos. Sci.* **1969**, *26*, 518–521.

(21) Schaller, R. C.; Fukuta, N. Ice Nucleation by Aerosol Particles: Experimental Studies Using a Wedge-Shaped Ice Thermal Diffusion Chamber. *J. Atmos. Sci.* **1979**, *36*, 1788–1802.

(22) Wang, B.; Knopf, D. A.; China, S.; Arey, B. W.; Harder, T. H.; Gilles, M. K.; Laskin, A. Direct Observation of Ice Nucleation Events on Individual Atmospheric Particles. *Phys. Chem. Chem. Phys.* **2016**, *18*, 29721–29731.

(23) Kiselev, A.; Bachmann, F.; Pedevilla, P.; Cox, S. J.; Michaelides, A.; Gerthsen, D.; Leisner, T. Active Sites in Heterogeneous Ice Nucleation—the Example of K-Rich Feldspars. *Science* **2017**, *355,* 367–371.

(24) Nath, S.; Boreyko, J. B. On Localized Vapor Pressure Gradients Governing Condensation and Frost Phenomena. *Langmuir* **2016**, *32*, 8350–8365.

(25) Eberle, P.; Tiwari, M. K.; Maitra, T.; Poulikakos, D. Rational Nanostructuring of Surfaces for Extraordinary Icephobicity. *Nanoscale* **2014**, *6*, 4874–4881.

(26) Campbell, J. M.; Meldrum, F. C.; Christenson, H. K. Is Ice Nucleation from Supercooled Water Insensitive to Surface Roughness? *J. Phys. Chem. C* **2015**, *119*, 1164–1169.

(27) Higuchi, K.; Noda, S. Selected Area Diamond Deposition by Control of the Nucleation Sites. *Diam. Relat. Mater.* **1992**, *1*, 220–229.

(28) Varga, M.; Vojs, M.; Marton, M.; Michalíková, L.; Veselý, M.; Redhammer, R.; Michalka, M. Diamond Thin Film Nucleation on Silicon by Ultrasonication in Various Mixtures. *Vacuum* **2012**, *86*, 681–683.

(29) Holbrough, J. L.; Campbell, J. M.; Meldrum, F. C.; Christenson, H. K. Topographical Control of Crystal Nucleation. *Cryst. Growth Des.* **2011**, *12*, 750–755.

(30) Campbell, J. M.; Meldrum, F. C.; Christenson, H. K. Observing the Formation of Ice and Organic Crystals in Active Sites. *Proc. Natl. Acad. Sci.* **2017**, *114*, 810–815.

(31) Na, B.; Webb, R. L. A Fundamental Understanding of Factors Affecting Frost Nucleation. *Int. J. Heat Mass Transf.* **2003**, *46*, 3797–3808.

(32) Marcolli, C.; Nagare, B.; Welti, A.; Lohmann, U. Ice Nucleation Efficiency of AgI: Review and New Insights. *Atmos. Chem. Phys.* **2016**, *16*, 8915–8937.

(33) Brumberg, A.; Hammonds, K.; Baker, I.; Backus, E. H. G.; Bisson, P. J.; Bonn, M.; Daghlian, C. P.; Mezger, M. D.; Shultz, M. J. Single-Crystal Ih Ice Surfaces Unveil Connection between Macroscopic and Molecular Structure. *Proc. Natl. Acad. Sci.* **2017**, *114*, 5349–5354.

(34) Shultz, M. J.; Bisson, P. J.; Brumberg, A. Best Face Forward: Crystal-Face Competition at the Ice-Water Interface. *J. Phys. Chem. B* **2014**, *118*, 7972–7980.

(35) Jähnert, S.; Vaca Chávez, F.; Schaumann, G. E.; Schreiber, A.; Schönhoff, M.; Findenegg, G. H. Melting and Freezing of Water in Cylindrical Silica Nanopores. *Phys. Chem. Chem. Phys.* **2008**, *10*, 6039–6051.

(36) Fitzner, M.; Sosso, G. C.; Cox, S. J.; Michaelides, A. The Many Faces of Heterogeneous Ice Nucleation: Interplay between Surface Morphology and Hydrophobicity. *J. Am. Chem. Soc.* **2015**, *137*, 13658–13669.

(37) Fukuta, N. Activation of Atmospheric Particles as Ice Nuclei in Cold and Dry Air. *J. Atmos. Sci.*

**1966**, *23*, 741–750.

(38) Fletcher, N. H. Active Sites and Ice Crystal Nucleation. *J. Atmos. Sci.* **1969**, *26*, 1266–1271.

(39) Christenson, H. K. Two-Step Crystal Nucleation *via* Capillary Condensation. *CrystEngComm* **2013**, *15*, 2030–2039.

(40) Marcolli, C. Deposition Nucleation Viewed as Homogeneous or Immersion Freezing in Pores and Cavities. *Atmos. Chem. Phys.* **2014**, *14*, 2071–2104.

(41) Kovács, T.; Meldrum, F. C.; Christenson, H. K. Crystal Nucleation without Supersaturation. *J. Phys. Chem. Lett.* **2012**, *3*, 1602–1606.

(42) Schreiber, A.; Ketelsen, I.; Findenegg, G. H. Melting and Freezing of Water in Ordered Mesoporous Silica Materials. *Phys. Chem. Chem. Phys.* **2001**, *3*, 1185–1195.

(43) Moore, E. B.; de la Llave, E.; Welke, K.; Scherlis, D. A.; Molinero, V. Freezing, Melting and Structure of Ice in a Hydrophilic Nanopore. *Phys. Chem. Chem. Phys.* **2010**, *12*, 4124–4134.

(44) Good, R. J. Contact Angle, Wetting, and Adhesion: A Critical Review. *J. Adhes. Sci. Technol.* **1992**, *12*, 1269–1302.

(45) O'Neill, G. A.; Westwater, J. W. Dropwise Condensation of Steam on Electroplated Silver Surfaces. *Int. J. Heat Mass Transf.* **1984**, *27*, 1539–1549.

(46) Ickes, L.; Welti, A.; Hoose, C.; Lohmann, U. Classical Nucleation Theory of Homogeneous Freezing of Water: Thermodynamic and Kinetic Parameters. *Phys. Chem. Chem. Phys.* **2015**, *17*, 5514–5537.

(47) Subramanyam, S. B.; Kondrashov, V.; Rühe, J.; Varanasi, K. K. Low Ice Adhesion on Nano-Textured Superhydrophobic Surfaces under Supersaturated Conditions. *ACS Appl. Mater. Interfaces* **2016**, *8*, 12583–12587.

(48) Davis, A.; Yeong, Y. H.; Steele, A.; Bayer, I. S.; Loth, E. Superhydrophobic Nanocomposite Surface Topography and Ice Adhesion. *ACS Appl. Mater. Interfaces* **2014**, *6*, 9272–9279.

(49) Murphy, D. M.; Koop, T. Review of the Vapour Pressures of Ice and Supercooled Water for Atmospheric Applications. *Q. J. R. Meteorol. Soc.* **2005**, *131*, 1539–1565.

# Figures

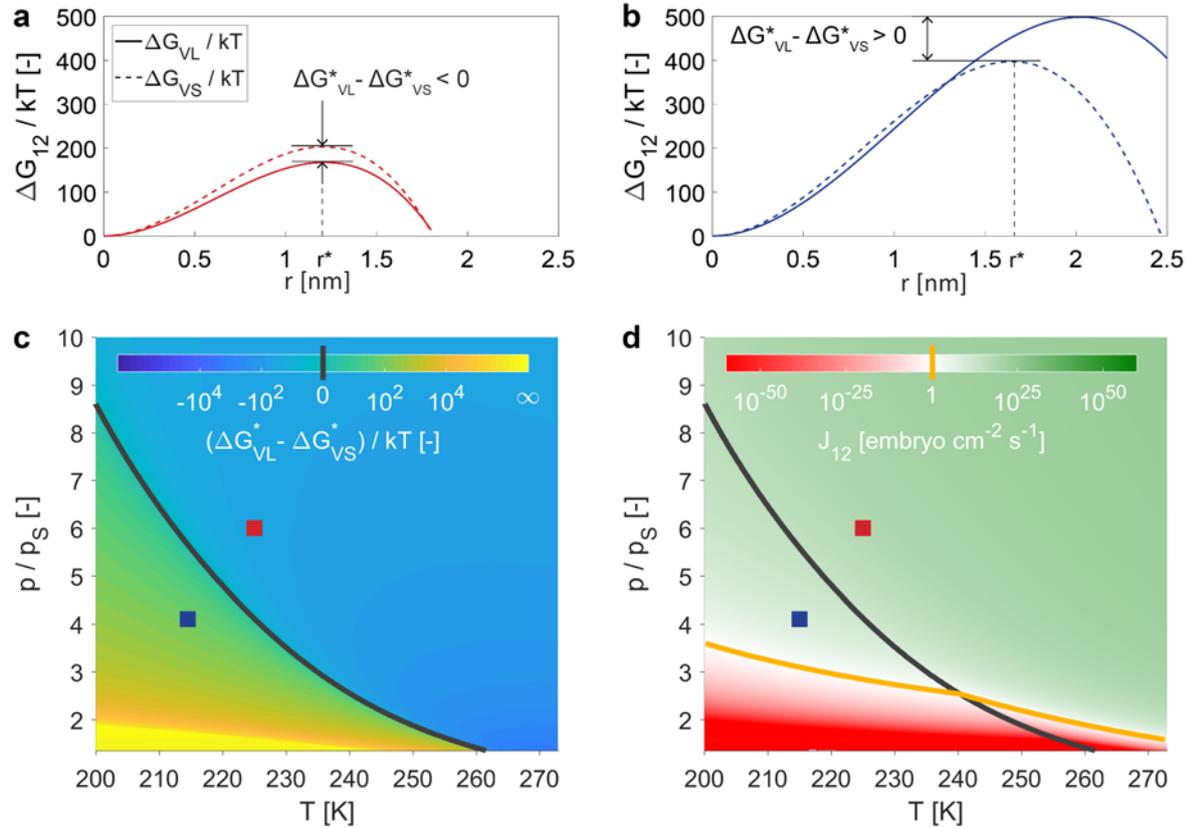

**Figure 1. Theoretical condensation-desublimation nucleation transition.** The difference between critical Gibbs free energy of homogeneous nucleation for condensation and desublimation ($\Delta G_{VL}^* - \Delta G_{VS}^*$) changes as a function of temperature ($T$) and water vapor pressure ($p$), which can cause a switch in energetic favorability for pathway to frost nucleation. (a, b) Plots showing normalized Gibbs free energy ($\Delta G_{12}/kT$) as a function of the nascent embryo radius ($r$). Plots exemplify two conditions of (a) $T$ = 225K, $p$ = 0.3 mbar, in which case $\Delta G_{VL}^* < \Delta G_{VS}^*$, indicating condensation nucleation is favorable and (b) $T$ = 215K, $p$ = 0.05 mbar, in which case $\Delta G_{VL}^* > \Delta G_{VS}^*$, indicating desublimation nucleation is favorable. Dotted lines plot $\Delta G_{VS}/kT$, solid lines plot $\Delta G_{VL}/kT$. (c) Surface plot, $(\Delta G_{VL}^* - \Delta G_{VS}^*)/kT$ vs. $p/p_S$ vs. $T$, indicating conditions where condensation ($(\Delta G_{VL}^* - \Delta G_{VS}^*)/kT < 0$) or desublimation ($(\Delta G_{VL}^* - \Delta G_{VS}^*)/kT > 0$) nucleation is energetically favorable. Red and blue squares indicate conditions illustrated in (a) $T$ = 225K, $p$ = 0.3 mbar and (b) $T$ = 215K, $p$ = 0.05 mbar respectively. Solid black line indicates zero point ($(\Delta G_{VL}^* - \Delta G_{VS}^*)/kT = 0$, equal favorability). (d) Surface plot for heterogeneous nucleation rate ($J_{12}$) vs. $p/p_S$ vs. $T$, indicating conditions on a surface ($\theta_{VL} = \theta_{VS}$ = 75° and $R = \infty$). The green area indicates that $J_{12}$ is large enough to observe nucleation on a reasonable time and area scale ($J_{12} > 1$ embryo cm$^{-2}$ s$^{-1}$) and the red area indicates that $J_{12}$ is too small to result in observed nucleation on a reasonable time and area scale ($J_{12} < 1$ embryo cm$^{-2}$ s$^{-1}$). Solid black line indicates $(\Delta G_{VL}^* - \Delta G_{VS}^*)/kT = 0$, equal nucleation pathway favorability for reference. Orange line indicates $J_{12} = 1$ embryo cm$^{-2}$ s$^{-1}$. Red and blue

squares indicate conditions illustrated in (a) $T$ = 225K, $p$ = 0.3 mbar and (b) $T$ = 215K, $p$ = 0.05 mbar respectively.

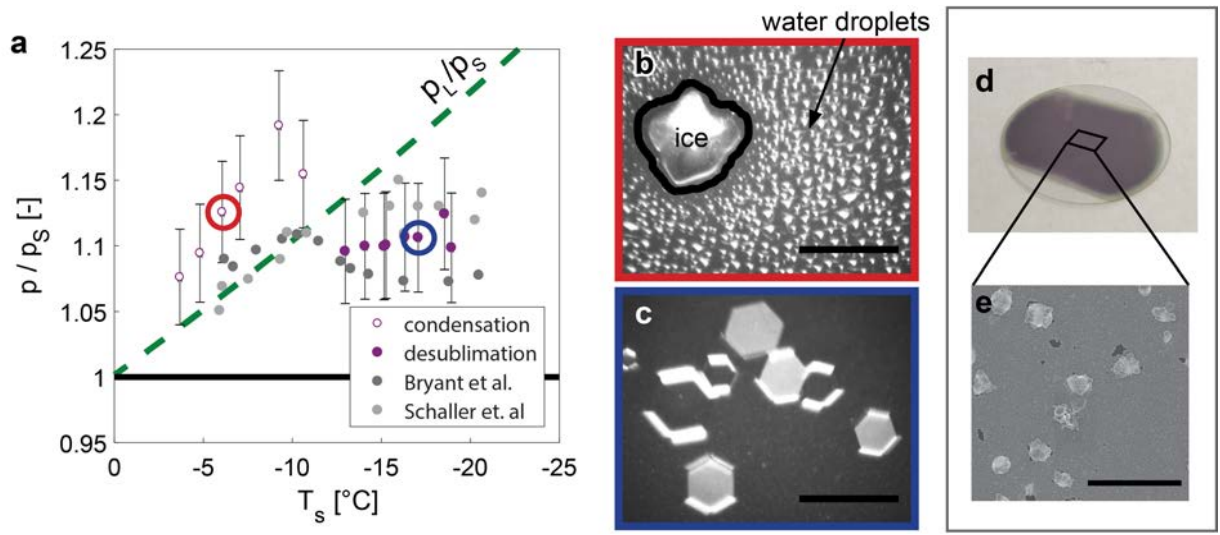

**Figure 2. Experimental condensation-desublimation nucleation transition.** The heterogeneous transition can be experimentally observed by plotting $T_s$ vs. $p/p_S$ at the onset of nucleation for a given surface and observing the nucleation mode. (a) Experimentally observed results for AgI (purple data points) correspond well to experiments done on AgI by Bryant[19] (dark grey data points) and Schaller[21] (light grey data points). Here, desublimation is observed (filled purple data points) when $p/p_S < p_L/p_S$ (green dashed line) and $p/p_S > 1$ (black solid line). Condensation is observed (unfilled purple data points) when $p/p_S > p_L/p_S$. AgI exhibits a condensation-desublimation nucleation transition at approximately $T_s$ = -12°C. Optical micrographs show examples of (b) condensation and (c) desublimation on AgI; circled data points in **a** correspond to condensation (red) and desublimation (blue). SEM micrograph (e) of AgI sample (d) used to conduct experiments shows inhomogeneous topography comprised of AgI crystals of the order of 0.5 μm. (b, c) Scale bar: 200 μm. (e) Scale bar: 2 μm.

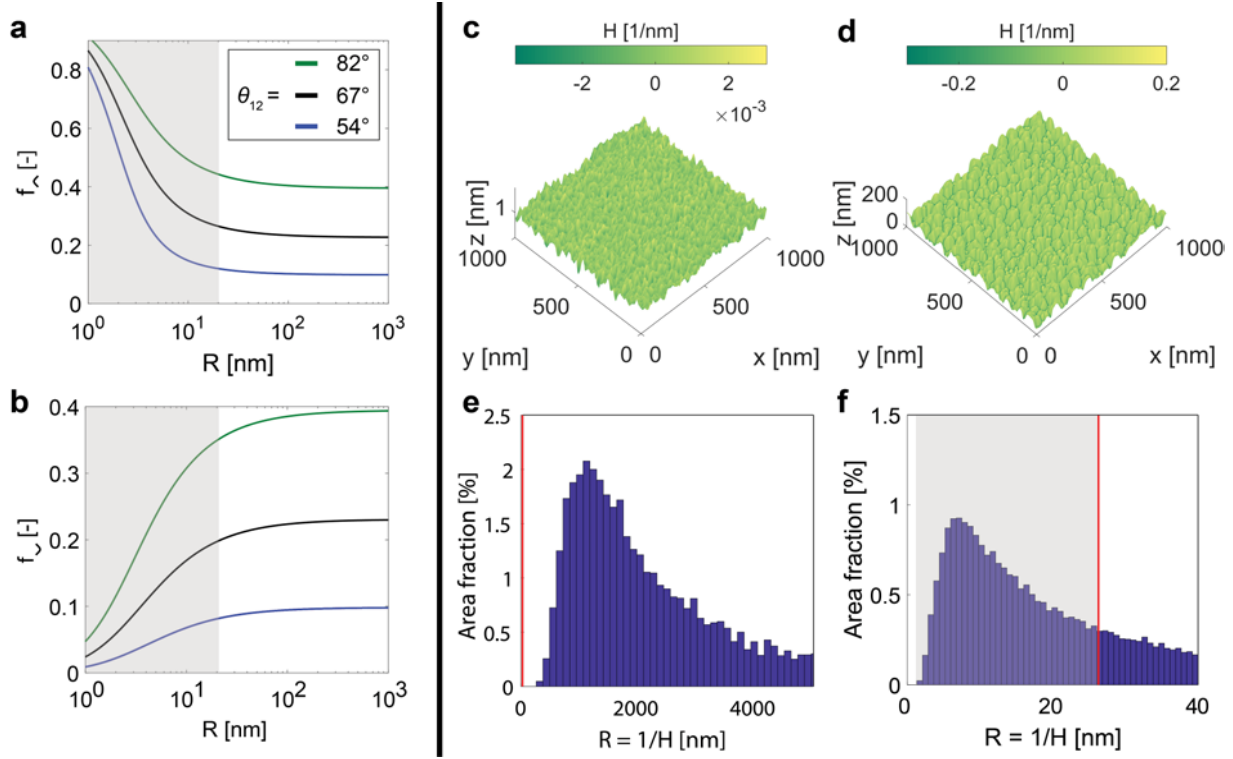

**Figure 3. Nanotexture can theoretically change the heterogeneous nucleation barrier to a varying degree dependent on $\theta_{12}$ and $r^*_{vs}$.** The change in $f(\theta_{12}, R)$ for both (a) bumps and (b) pits under conditions $T_\infty$ = -47.9 – -55.2°C (corresponding to $p_S$ = 0.05 to 0.02 mbar), $p/p_S$ = 1.45, for $\theta_{12}$ = 54° (solid blue line), 67° (solid black line), and 81° (solid green line). (c, d) AFM micrographs of smooth silicon (RMS roughness = 0.94 nm) and nanotextured silicon surface (RMS roughness = 61.8 nm), respectively, used in experiments (AFM tip radius 2 nm); the mean curvature $H$ is overlaid on height profile as a surface plot. (e, f) Histograms showing the surface area fraction as a percentage of the total surface occupied by different magnitudes of $R$. The solid red line indicates $R = 10\, r^*_{12}$. The smooth silicon surfaces (e) contain no $R \leq 10\, r^*_{12}$, while the nanotexture silicon (f) has 18% surface coverage of $R \leq 10\, r^*_{12}$. Transparent grey area in **a** and **b** indicates $R < 10\, r^*_{12}$ corresponding to the same $R$ also indicated by transparent grey area in **f**, illustrating that the engineered nanotexture quantified in **d** should have an influence on the heterogeneous nucleation behavior.

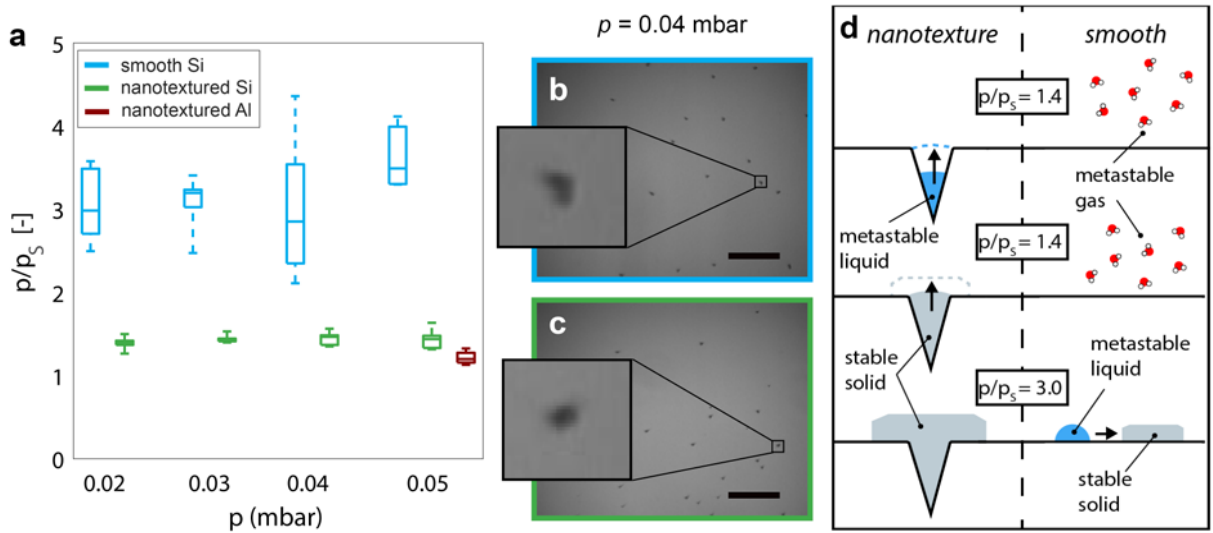

**Figure 4. Experimentally measured effect of nanotexture on heterogeneous nucleation.** (a) Box plot showing experimental measured $p/p_S$ at first observation of nucleation on smooth silicon (blue), nanotextured silicon (green), and nanotextured aluminum (red) vs. $p$ corresponding to limit of experimental chamber ($T_\infty$ = -47.9 – -55.2°C). Boxes represent the experimental spread for nine individual experiments (n=9). The nine experiments were done on three different samples of each sample type. The results indicate that it is statically more likely for nucleation to occur on nanotextured silicon than smooth silicon at a lower $p/p_S$. Nanotextured aluminum also indicates earlier onset of nucleation, comparable to that of nanotextured silicon. Crystal growth observed on (b) smooth silicon and (c) nanotextured silicon at $p$ = 0.04 mbar 30 seconds after first observed nucleation. (b, c) Scale bar: 200 μm. (d) Mechanism to explain the earlier observed nucleation on nanotexture. At $p/p_S$ = 1.4, metastable liquid can nucleate on the nanotextured silicon due to the Kelvin effect, while metastable vapor is still present above the smooth silicon. Due to the considerable supercooling (below the homogenous nucleation limit), the small volume of metastable liquid freezes in the pores and is able to grow to be macroscopically visible, while the vapor over the smooth silicon remains metastable. At $p/p_S$ = 3.0, metastable liquid can finally nucleate on the smooth silicon and due to the considerable supercooling, freezes.

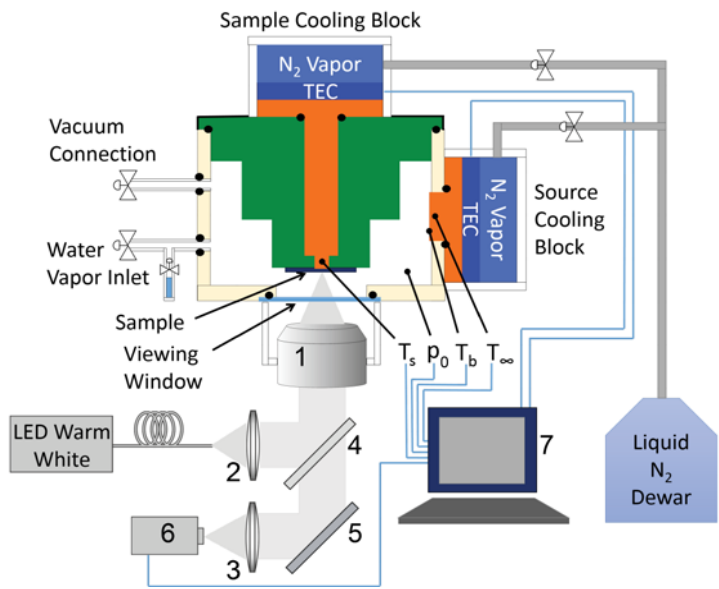

**Figure 5. Illustration of experimental setup.** Experimental setup consists of experimental chamber with connection to vacuum, water vapor inlet, and sample and source cooling block; imaging system with LED warm white light source, (1) focusing objective, (2) collimating lens, (3) focusing lens, (4) beam splitter, (5) mirror, and (6) CMOS camera; computer with data acquisition system recording sample temperature ($T_s$), chamber pressure ($p_0$), source cooling block temperature inside chamber ($T_b$) and in copper block ($T_\infty$), and camera image and individual PID control of both thermoelectric coolers (TECs); and liquid nitrogen dewar.

# Supporting Information

## Desublimation Frosting on Nanoengineered Surfaces

*Christopher Walker, Sebastian Lerch, Matthias Reininger, Hadi Eghlidi, Athanasios Milionis, Thomas M. Schutzius\*, Dimos Poulikakos\**

Laboratory of Thermodynamics in Emerging Technologies, Department of Mechanical and Process Engineering, ETH Zurich, Sonneggstrasse 3, CH-8092 Zurich, Switzerland

**Keywords: frosting, icing, condensation, desublimation, surface engineering, icephobic**

\*To whom correspondence should be addressed.

Prof. Dimos Poulikakos
ETH Zurich
Laboratory of Thermodynamics in Emerging Technologies
Sonneggstrasse 3, ML J 36
CH-8092 Zurich
Switzerland
Phone: +41 44 632 27 38
Fax:+41 44 632 11 76
dpoulikakos@ethz.ch

Dr. Thomas M. Schutzius
ETH Zurich
Laboratory of Thermodynamics in Emerging Technologies
Sonneggstrasse 3, ML J 27.2
CH-8092 Zurich
Switzerland
Phone: +41 44 632 46 04
thomschu@ethz.ch

## S1 Critical Gibbs Free Energy

It is instructive to consider the change in Gibbs free energy necessary for homogeneous (in the absence of a foreign surface) nucleation ($\Delta G_{12}$). We can define it as $\Delta G_{12} = n_2(\mu_2 - \mu_1)v + A\gamma_{12}$, where $n_2$ is the number of molecules per unit volume of the daughter phase, $\mu_1$ and $\mu_2$ are the chemical potentials of the parent and daughter phases, respectively, $v$ is the nascent embryo volume, $A$ is the nascent embryo surface area, and $\gamma_{12}$ is the interfacial energy between the parent and daughter phase. We can define the difference in chemical potential between the daughter and the parent phase as $\mu_2 - \mu_1 = -kT \ln(p/p_2)$, where $p$ and $p_2$ are the water vapor pressure in the environment and saturated vapor pressure over a plane surface of the daughter phase, respectively, at temperature $T$, and $k$ is the Boltzmann constant. Assuming that $v = (4/3)\pi r^3 \alpha$ and $A = 4\pi r^2 \beta$, where $\alpha$ and $\beta$ are numerical factors both greater than unity used to adjust for non-spherical embryos shapes, and $r$ is the embryo radius, we obtain an equation for $\Delta G_{12}$ as a function of $r$,

$$\Delta G_{12} = -4/3\,\pi r^3 \alpha n_2 kT \ln(p/p_2) + 4\pi r^2 \beta \gamma_{12} \,. \tag{S6}$$

From Eq. (S1), we see that initially as $r$ increases so does $\Delta G_{12}$ (*ceteris paribus*); however, after the embryo has reached a critical size, $r^*$, $\Delta G_{12}$ begins to decrease with a further increasing $r$; therefore, growth of a stable nascent embryo is spontaneous. We can define $r_{12}^*$ for condensation and desublimation by setting $\partial \Delta G_{12}/\partial r = 0$, which yields $r_{12}^* = 2\gamma_{12}/(n_2 kT \ln(p/p_2))$. Substitution of $r_{12}^*$ for $r$ into Eq. (S1) yields

$$\Delta G_{12}^* = 16\pi \gamma_{12}^3 \xi / 3\left[n_2 kT \ln(p/p_2)\right]^2 \tag{S7}$$

where $\xi = \beta^3 / \alpha^2$.[1] Allowing the subscripts $V$, $L$, and $S$ take the place of parent phase (1) and daughter phase (2) explicitly as the vapor phase, liquid phase, and solid phase, respectively, the corresponding critical free energies for condensation and desublimation are

$$\Delta G_{VL}^* = 16\pi \gamma_{VL}^3 \xi / 3 \left[ n_L kT \ln \left( p / p_L \right) \right]^2, \tag{S8}$$

$$\Delta G_{VS}^* = 16\pi \gamma_{VS}^3 \xi / 3 \left[ n_S kT \ln \left( p / p_S \right) \right]^2, \tag{S9}$$

respectively.

## S2 Nucleation Rate

The rate at which nucleation of daughter (2) in parent (1) phase occurs is calculated using an Arrhenius equation, $J_{12} = J \exp(-\Delta G^*_{12}/kT)$, where $J$ is a pre-exponential factor, which depends on the rate at which vapor molecules strike a unit area of the embryo. This rate is given from kinetic gas theory as $p/(2\pi mkT)^{1/2}$, where $m$ is the mass of a water molecule. Therefore, in a unit volume of vapor, the rate of formation of the daughter embryos that can grow spontaneously is given as[1]

$$J_{12} = \frac{p 4\pi r^{*2}_{12} \beta n_1}{(2\pi mkT)^{1/2}} \exp\left(-\frac{\Delta G^*_{12}}{kT}\right). \tag{S10}$$

To compare the rate of stable embryo formation for condensation and desublimation, we can first study the pre-exponential term in Eq. (S5). For condensation and desublimation, the pre-exponential terms only differ by a factor $\beta$ and in this case are equal to within a factor of two as the other variables do not change.[1] Therefore it is commonly assumed that the pre-exponential term for condensation and desublimation are the same and given as $J$ = 1e29 $m^{-3} s^{-1}$.[2] This allows the nucleation rates for condensation and desublimation to be given as

$$J_{VL} = J \exp(-\Delta G^*_{VL}/kT), \tag{S11}$$

$$J_{VS} = J \exp(-\Delta G^*_{VS}/kT), \tag{S12}$$

respectively.

## S3 Homogeneous and Heterogeneous Nucleation: Geometric Factor

When considering heterogeneous nucleation, $\Delta G_{12}^*$ (see section "Critical Gibbs Free Energy") is modified by a function $f(\theta_{12}, R) \leq 1$, defined as the ratio of heterogeneous to homogeneous nucleation free energy barrier, which depends on the inherent contact angle of the daughter phase on the surface in the parent phase ($\theta_{12}$) and the surface's roughness radius of curvature ($R$) at the location of the nascent daughter phase embryo. If we only consider the influence of $\theta_{12}$ it becomes[1,3]

$$f(\theta_{12}) = 1/4\,(2 - 3\cos\theta_{12} + \cos^3\theta_{12}). \tag{S13}$$

$f$ becomes considerably more complicated when also considering the radius of curvature of the nanotexture ($R$). When considering concave radii of curvature (negative, nanopits) we obtain[4]

$$f(m,R) = \frac{1}{2}\left\{1 - \left(\frac{1+mx}{g_c}\right)^3 - x^3\left[2 - 3\left(\frac{x+m}{g_c}\right) + \left(\frac{x+m}{g_c}\right)^3\right] + 3mx^2\left(\frac{x+m}{g_c} - 1\right)\right\} \tag{S14}$$

where $m = \cos\theta_{12}$, $g_c = (1 + 2mx + x^2)^{1/2}$, and $x = R/r_{12}^*$. For convex radii of curvature (positive, nanobumps) we obtain

$$f(m,R) = \frac{1}{2}\left\{1 + \left(\frac{1-mx}{g_c}\right)^3 + x^3\left[2 - 3\left(\frac{x-m}{g_c}\right) + \left(\frac{x-m}{g_c}\right)^3\right] + 3mx^2\left(\frac{x-m}{g_c} - 1\right)\right\} \tag{S15}$$

where $m = \cos\theta_{12}$, $g_c = (1 - 2mx + x^2)^{1/2}$, and $x = R/r_{12}^*$. To simplify our discussion in the main body of the paper, we define $f$ directly as a function of the substrate properties, $\theta_{12}$ and $R$.

## S4 Ice-vapor-substrate Contact Angle

Conventional goniometers are able to measure $\theta_{VL}^*$ by simply observing the equilibrium angle formed when a liquid is brought into contact with a solid in a vapor environment. Unfortunately, a similar approach is not possible to measure $\theta_{VS}^*$, due to the obvious reason that ice does not readily change its shape to form the lowest energy state when brought into contact with a solid. We propose a method to experimentally measure $\theta_{VS}^*$ based upon the results of nucleation experiments. The method works based upon the observation of desublimation nucleation on the surface of interest and uses the experimentally obtained values of $p/p_S$ values as nucleation is observed in the desublimation region. The calculation methodology is explained.

Using Eq. (4), from the main text, and substituting conventional values[1–3] for $J_{VS}$ = 1 embryo cm$^{-2}$s$^{-1}$ and $J$ = 1e29 m$^{-3}$s$^{-1}$ we can calculate that $\Delta G_{VS}^*/kT$ = 57.5 will result in a nucleation rate within a reasonable amount of time (1 s) and space (1 cm²). Modifying Eq. (2), from the main text, by dividing both sides by $kT$, we substitute our experimentally obtained $p/p_S$ values when nucleation is observed in the desublimation region, to calculate an initial $\Delta G_{VS}^*/kT$ based on our experimental results. This initial calculation disregards any effect of heterogeneous nucleation and therefore its corresponding $f(\theta_{VS}, R)$, which is naturally incorrect, but provides us the means to work backwards to calculate $f(\theta_{VS}, R)$.

Reminding ourselves of the calculation for heterogeneous nucleation as a function of homogeneous nucleation barrier: $\dfrac{\Delta G_{12}^*}{kT} f(\theta_{12}, R)$, we know that this value must be below 57.5 due to our experimental observations of seeing desublimation nucleation at a certain $p/p_S$ and $T$ used to calculate $\Delta G_{VS}^*/kT$. We calculate $f(\theta_{VS}, R)$ to be $\Delta G_{12}^*/57.5kT$, which then can be used to calculate $\theta_{12}$ through use of Eq. (S8). It should be noted that we assume here that $R$ is large enough and therefore does not play a role in promoting or discouraging nucleation.

## S5 Wettability of AgI

The approach explained in the previous section (see section "Ice-vapor-substrate Contact Angle") was used to calculate an effective vapor solid contact angle for silver iodide (AgI). Due to the inhomogeneous surface texture of the AgI samples used, we cannot claim with certainty that we have calculated $\theta_{VS}$, but rather an effective $\theta_{VS}$, which we term $\theta_{VS}^*$, in which surface nanotexture may play a role. An analogous approach to the method explained earlier can be employed to calculate $\theta_{VL}^*$, by using instead our experimentally measured $p/p_L$. Values for $p/p_L$ and $p/p_S$ were obtained from our experimental results, plotted in Figure 2a, by averaging the values in the condensation and desublimation regions, resulting in $p/p_L = 1.05$ and $p/p_S = 1.10$ respectively. Substituting these values into Eqs. (1) and (2) from the main text, we obtain $\Delta G_{VL}^*/kT = 6.3 \times 10^4$ and $\Delta G_{VS}^*/kT = 4.2 \times 10^4$ at $T_s$ = -12 °C. Due to the influence of the AgI surface, the free energy barrier to nucleation is reduced to $\Delta G_{12}^*/kT = 57.5$ (see section "Ice-vapor-substrate Contact Angle") due to $f(\theta_{12}, R)$ for condensation ($f_{VL}$) and desublimation ($f_{VS}$) respectively. We calculate $f_{VL} = 9.1 \times 10^{-4}$ and $f_{VS} = 0.0014$. We can disregard the influence of $R$ on $f_{12}$ by solving for $\theta_{12}^*$ instead of $\theta_{12}$ using Eq. (S8). We obtain the values $\theta_{VL}^* = 15°$ and $\theta_{VS}^* = 17°$, which correspond well to previously calculated literature values of $\theta_{VL}^* = 10°$ and $\theta_{VS}^* = 16°$.[5]

## S6 Wettability of Silicon

Due to the variance of $p/p_S$ upon observation of nucleation for smooth silicon at constant temperatures, we were not able to obtain a clear condensation-desublimation nucleation transition that would be necessary to determine $\theta_{VS}$ with the help of experiments, as done with AgI. Therefore we resorted to a range of estimations for this value to use in the analysis of the effect of nanotexture. In order to do this we defined $\theta_{VL}$ for our silicon samples using literature values. By estimating that a certain layer thickness of 8 – 10 angstroms of native oxide had grown on our silicon samples during two minute oxygen plasma and exposure to air over a period of less than seven days,[6] we could calculate the influence of the native oxide layer on the contact angle of silicon,[7] and estimate a contact angle to be 60 – 65°. $\theta_{VS}$ can be related to $\theta_{VL}$ by combining the respective Young-Dupre equations for a water droplet embryo in vapor and an ice crystal embryo in vapor to obtain

$$\cos\theta_{VS} = (\gamma_{VL}/\gamma_{VS})\cos\theta_{VL} + (\gamma_{LA} - \gamma_{SA})/\gamma_{VS} \tag{S16}$$

where $\gamma_{12}$ refers to the interfacial energy between two interfaces, with subscripts $V$, $L$, $S$, and $A$ refer to the interfaces of vapor, liquid (water), solid (ice), and sample surface respectively. Considering our estimate for $\theta_{VL}$, and existing tabulated values for $\gamma_{VL}$ and $\gamma_{VS}$ as a function of $T_s$, the single unknown in Eq. (S11) necessary to calculate a value for $\theta_{VS}$ is the difference in the surface energy of the sample surface to water and ice ($\gamma_{LA} - \gamma_{SA}$). For our calculations of $\theta_{VS}$ for silicon, we used $\gamma_{LA} - \gamma_{SA}$ =0.021 J m$^{-2}$ as an upper bound. This is a value calculated for AgI substituting the values of $\theta_{VS}^*$ and $\theta_{VL}^*$ into Eq. (S11). The particularly low surface energy of silver iodide and ice justifies this as an upper bound. We use $\gamma_{LA} - \gamma_{SA}$ = -0.021 J m$^{-2}$ as a lower bound. For most surfaces, $\gamma_{LA} < \gamma_{SA}$ due to the smaller interfacial energy between a liquid and a solid when compared to a solid and a solid. Using Eq. (S11), both bounds of $\gamma_{LA} - \gamma_{SA}$ and an intermediate of $\gamma_{LA} - \gamma_{SA}$ = 0 J m$^{-2}$, and the previously calculated values for $\theta_{VL}$ of silicon, we obtain a range three plausible values for for $\theta_{VS}$ which are used in our analysis for the theoretical effect of nanotexture on the nucleation in Figure 3a and b.

## S7 Effect of Roughness on Nucleation Rate

It is important to understand how nanotexture globally influences the rate of nucleation ($J_{12}$) on a surface. Since nanotexture both enhances and diminishes the rate of nucleation through pits and bumps respectively, it is at first glance, unclear what the overall effect of nanotexture is, with respect to the global nucleation rate when compared to a smooth surface of the same composition. It turns out that due to the exponential dependence of $J_{12}$ on $f(\theta_{12}, R)$, the increase in $J_{12}$ caused by pits significantly outweighs the decrease in $J_{12}$ caused by bumps and that the overall $J_{12}$ for a nanotextured surface is increased significantly. This section demonstrates how we arrive at this outcome.

We begin by quantifying the heterogeneous nucleation rate for a smooth surface ($J_S$) by adapting Eq. (S5),

$$J_S = J \exp\left(-f(\theta_{12})\Delta G_{12}^*/kT\right) \tag{S17}$$

where we have changed the pre-exponential factor to a constant ($J$) for simplicity and included the heterogeneous function, $f(\theta_{12})$, for $R \to \infty$. In order to quantify $J_{12}$ for a nanotextured surface we make the assumption that all pits and bumps have the same $R$ and that there is an even distribution of pits and bumps across the whole surface. In other words the surface coverage of bumps ($\phi_b$) and pits ($\phi_b$) are both 0.5. We can then quantify the heterogeneous nucleation rate for our idealized nanotextured surface ($J_N$) by taking a weighted average of the respective $J_{12}$ for bumps and pits,

$$J_N = J_0 \left( \phi_b \exp\left(\left(-f_b(\theta_{12}, R)\Delta G_{12}^*\right)/kT\right) + \phi_p \exp\left(\left(-f_p(\theta_{12}, R)\Delta G_{12}^*\right)/kT\right) \right) \tag{S18}$$

where $f_b(\theta_{12}, R)$ and $f_p(\theta_{12}, R)$ are the homogeneous modifying functions for bumps and pits of $R$ respectively. Figures 3a and b indicate that $f(\theta_{12}, R)$ changes substantially when nanotexture is present. Since a small change in the exponential argument leads to an order of magnitude larger change in the result, the large increase in the first term of $J_N$ added to a small second term leads to $J_N \gg J_S$.

# References

(1) Hobbs, P. V. Ice Physics. Oxford University Press: Oxford, 1974; pp. 461–571.

(2) Na, B.; Webb, R. L. A Fundamental Understanding of Factors Affecting Frost Nucleation. *Int. J. Heat Mass Transf.* **2003**, *46*, 3797–3808.

(3) Pruppacher, H. R.; Klett, J. D. Microphysics of Clouds and Precipitation. In *Atmospheric and Oceanographic Science Library*; Mysak, L.A., Hamilton K., Eds.; Springer: Heidelberg, 1998.

(4) Fletcher, N. H. Size Effect in Heterogeneous Nucleation. *J. Chem. Phys.* **1958**, *29*, 572.

(5) Manton, M. J. Parameterization of Ice Nucleation on Insoluble Particles. *Tellus B* **1983**, *35 B*, 275–283.

(6) Morita, M.; Ohmi, T.; Hasegawa, E.; Kawakami, M.; Ohwada, M. Growth of Native Oxide on a Silicon Surface. *J. Appl. Phys.* **1990**, *68*, 1272–1281.

(7) Williams, R.; Goodman, A. M. Wetting of Thin Layers of SiO2 by Water. *Appl. Phys. Lett.* **1974**, *25*, 531–532.

(8) Shultz, M. J.; Bisson, P. J.; Brumberg, A. Best Face Forward: Crystal-Face Competition at the Ice-Water Interface. *J. Phys. Chem. B* **2014**, *118*, 7972–7980.

# Figures

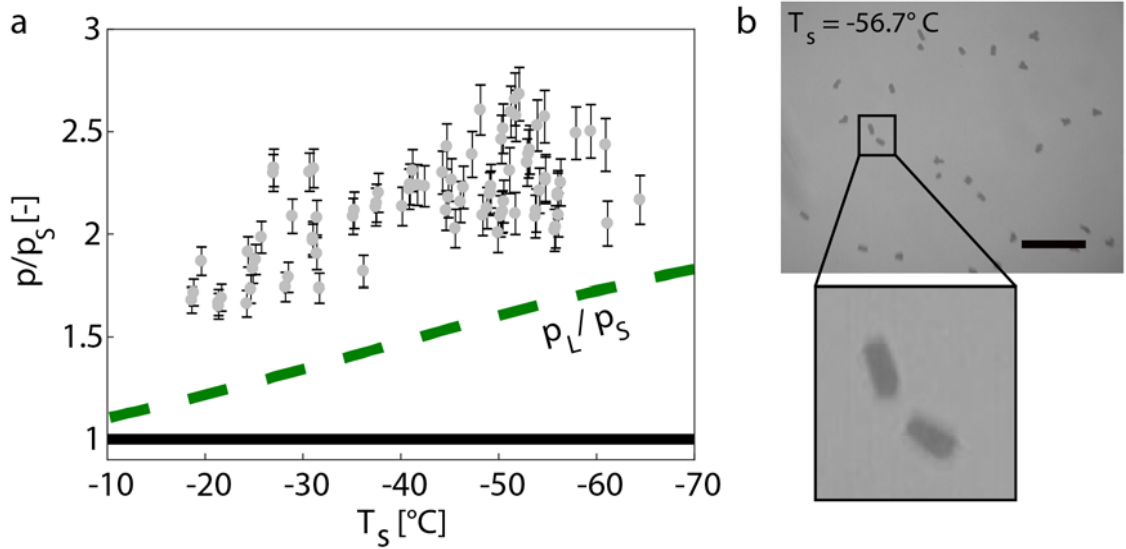

**Figure S1. Experimental nucleation conditions for silicon.** The heterogeneous condensation-desublimation nucleation transition for silicon was not be experimentally observed for 0°C < $T_s$ < -65°C. (a) Experimentally observed results for silicon (gray data points) all fall above $p/p_S > p_L/p_S$ (green dashed line). Even at extremely cold $T_s$ (< -60°C) there was no clear evidence for a transition to a desublimation regime similar to what was observed on AgI (see Figure 2a), providing evidence that condensation followed by freezing remains the energetically favorable pathway to frost for all $T_s$ used in this study. (b) Optical micrograph shows example of frost crystals grown at $T_s$ = -56.7 °C. Unlike the crystal shape observed in the desublimation region for AgI, characteristic of the basal face of ice 1h (see Figure 2c), the crystal shape observed on silicon is characteristic of the secondary prism face of ice 1h, providing further evidence of growth from the liquid-solid interface.[8] Scale bar: 200 μm.

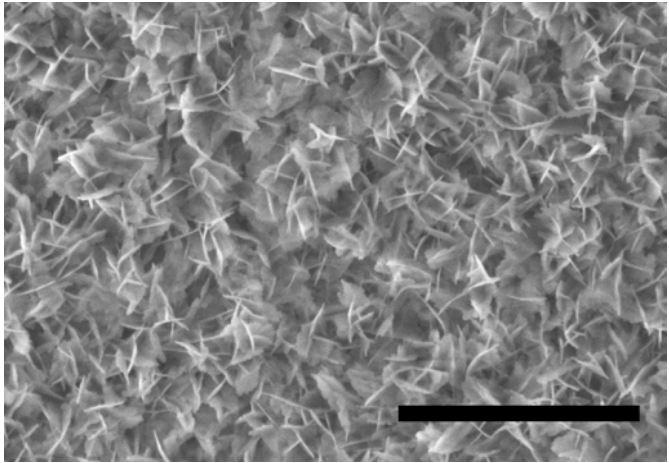

**Figure S2. SEM micrograph of boehmite nanostructures on aluminum.** Scale bar: 1μm.

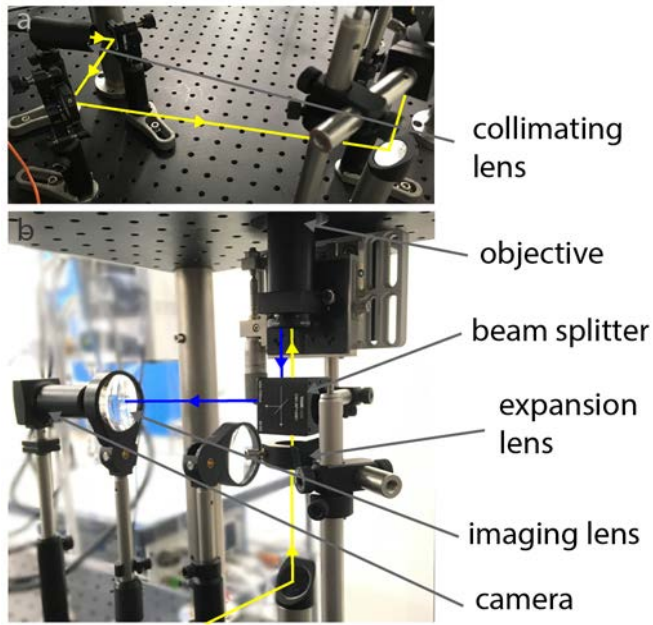

**Figure S3. Optical train used for nucleation observations.** The optical train used in the experiments consists of a reflected bright-field microscope that utilizes an LED warm white light. (a) The light from the LED source (yellow line) is collimated and aligned perpendicularly to the sample using a lens and a series of mirrors. (b) The light travels through an expansion lens (used to focus the light on the back focal plane of the objective), followed by a beam splitter (necessary to split the excitation and detection light paths), followed by an objective (used to illuminate the sample with a collimated beam and also collect the detected light). The light reflected from the sample (blue line) travels back through the same objective used for illumination and is reflected by the beam splitter to the imaging lens, which focuses the light onto an 8-bit CMOS camera chip, producing an image on the computer. This image is synchronized with the temperature and pressure measurements. The experimental chamber is held on the breadboard above the objective.